\documentclass[universe,article,accept,moreauthors,pdftex,a4paper]{mdpi} 

\usepackage{bbm}
\usepackage{amsfonts,amsmath,bm}
\usepackage{mathtools}

\newcommand{\dd}{\mathrm d}
\newcommand{\ex}{\mathrm e}
\newcommand{\GN}{G_\textsc{n}}
\newcommand{\Hu}{\mathcal{H}}
\newcommand{\Ka}{\mathcal{K}}
\newcommand{\dBB}{{\text{d}\textsc{bb}}}
\newcommand{\ud}{\mathrm{d}}

\renewcommand{\dot}[1] {\overset{\,_{\mbox{\Large .}}}{#1}}
\renewcommand{\ddot}[1] {\overset{\,_{\mbox{\Large ..}}}{#1}}

\firstpage{1} 
\pubvolume{2}
\issuenum{3}
\articlenumber{0}
\pubyear{2022}
\copyrightyear{2022}
\history{}

\pdfoutput=1

\Title{Clocks and Trajectories in Quantum Cosmology}

\Author{Przemys{\l}aw Ma{\l}kiewicz $^1$\orcidA{}, Patrick
Peter $^{2,}$*\orcidB{} and Sandro Dias Pinto
Vitenti $^{3,4}$\orcidC{}}

\AuthorNames{Przemys{\l}aw Ma{\l}kiewicz, Patrick Peter and
S. D. P. Vitenti}
\address{%
$^{1}$ \quad National Centre for Nuclear Research, Pasteura 7,
02-093 Warszawa, Poland; Przemyslaw.Malkiewicz@ncbj.gov.pl
\\ 
$^{2}$ \quad ${\cal
G}\mathbb{R}\varepsilon\mathbb{C}{\cal O}$---Institut
d'Astrophysique de Paris, CNRS \& Sorbonne Universit\'e, UMR
7095 98 bis Boulevard Arago, 75014 Paris, France\\ 
$^{3}$ \quad
Departamento de F\'{i}sica, Universidade Estadual de
Londrina, Rod. Celso Garcia Cid, Km 380, \mbox{Londrina 86057-970},
PR, Brazil; vitenti@uel.br\\ 
$^{4}$ \quad Instituto de F\'{\i}sica,
Universidade de Bras\'{\i}lia---UnB, Campus Universit\'ario
Darcy Ribeiro-Asa Norte Sala BT 297-ICC-Centro, 70919-970
Bras\'{\i}lia, Brazil}

\corres{Correspondence: peter@iap.fr}

\simplesumm{We clarify the question of clock transformations 
and trajectories in quantum cosmology in a vacuum Bianchi I
minisuperspace.}

\abstract{We consider a simple cosmological model consisting
of an empty Bianchi I Universe, whose Hamiltonian we
deparametrise to provide a natural clock variable. The model
thus effectively describes an isotropic universe with an
induced clock given by the shear. By quantising this model, we
obtain various different possible bouncing trajectories
(semiquantum expectation values on coherent states or
obtained by the de Broglie--Bohm formulation) and explicit
their clock dependence, specifically emphasising the question
of symmetry across the bounce.}


\begin{document}

\section{Introduction}

The problem of time in quantum
cosmology~\cite{Anderson:2017jij,Kiefer:2021zdq} is
well-known and, as of now, unsolved. It rests on the fact
that general relativity (GR) is a totally constrained
theory, and its canonically quantised counterpart can be
reduced to the Wheeler--DeWitt (WDW) equation $\Hu \Psi =0$,
which is a Schr\"odinger equation without time. Hence, dynamics is absent and, in a sense, meaningless in this framework.

A simple way to reintroduce dynamical properties into the
theory consists in deparametrisation, namely by making use
of the fact that there exists a constraint and using a
variable to serve as clock. Indeed, let us denote the
relevant canonical variables $\{ q^k\}$ and their associated
momenta $\{ p_k\}$, one has $\Hu \left( \{ q^k\}, \{ p_k\}
\right) \approx 0$ in the Dirac weak sense. Performing a
canonical transformation $\left( \{ q^k\}, \{ p_k\} \right)
\mapsto \left( \{ Q^a\}, \{ P_a\} \right)$ and assuming that
there exists a new variable $Q^\alpha$ such that the Poisson
bracket $\{ Q^\alpha ,\Hu \}_\textsc{p.b.}$ is unity, one
obtains $\dd Q^\alpha/\dd t = 1$, so that the variable
$Q^\alpha$ itself can be used as time; this is a classical
internal clock.

A simple and illustrative example consists in the
Hamiltonian $H_{\bm{x}y} = H_{\bm{x}} + H_y$ with arbitrary 
$H_{\bm{x}}$ for a set of variables $\bm{x}$ but
independent of the variable $y$, and $H_y = -\frac12
(\dot{y}^2 +y^2)$ represents a harmonic oscillator with
negative sign. The (local) canonical transformation $T=2 \arctan
(p_y/y)$ and $p_T = -\frac12(p_y^2+y^2)$ produces $H_y =
p_T$, leading to $\dot{p}_T = 0$ and $\dot{T}=1$, showing
that $T$ is a perfectly acceptable (local) clock variable for the
Hamiltonian $H_{\bm{x}}$.

Denoting $Q^\alpha \to t$ and its canonically conjugate
momentum $P_\alpha \to P_t$, one notes that since $\{
Q^\alpha ,\Hu \}_\textsc{p.b.} =1$, the total Hamiltonian
can be split into $\Hu = P_t + H$, where $H$ may depend on
$t$ but not on $P_t$. At the quantum level, it then suffices
to apply the Dirac operator prescription $p_t\mapsto
\hat{p}_t = - i\hbar \partial/\partial t$ to the original
time WDW equation without time to transform it into $i\hbar
\partial \Psi/\partial t = H\Psi$ and thus recover a
time-dependent Schr\"odinger equation. Although this
procedure is not always applicable for configurations in
superspace, restriction to a cosmological minisuperspace
often permits it.

The question that naturally comes to mind is whether a clock
thus defined is unique and what the effect of changing it
is. In what follows, we first discuss a simple cosmological
model based on a homogeneous but anisotropic Bianchi I
metric in Section~\ref{sec_BI} in which we obtain a clock
provided by the shear; this yields a simple free-particle
Hamiltonian in which we introduce an affine quantisation
procedure (Section~\ref{sec_affine}) to account for the
restriction that the scale factor is positive definite.
Section~\ref{sec_clock} is dedicated to exploring in detail
the clock transformations relevant to our quantised model,
and we discuss the associated trajectories in
Section~\ref{sec_traj} before wrapping up our findings and
concluding.

\section{Classical Bianchi I Model} \label{sec_BI}

We begin by assuming a homogeneous and anisotropic Bianchi type I
metric
\begin{equation}
\dd s^2 = -N^2\dd \tau^2 +
\underbrace{\ex^{2(\beta_0 + \beta_+ + \sqrt{3}\beta_-)}}_{a^2_1}
\left(\dd  x^1\right)^2
+ \underbrace{\ex^{2(\beta_0 + \beta_+ - \sqrt{3}\beta_-)}}_{a^2_2}
\left(\dd x^2\right)^2
+ \underbrace{\ex^{2(\beta_0 - 2\beta_+)}}_{a^2_3}
\left(\dd x^3\right)^2,
\label{BI}
\end{equation}
thereby defining the scale factors $a_i$ and the lapse $N$. Classically,
in order to ensure the required symmetries, all these functions are
assumed to depend on time $\tau$ only. For the metric \eqref{BI},
the usual Einstein--Hilbert action then reduces to

\begin{equation}
\mathcal{S}_\textsc{eh}=\frac{1}{16\pi\GN}\int\sqrt{-g} R\dd^4 x 
= \frac{3}{8\pi\GN} \underbrace{\int
  \sqrt{\gamma}\dd^3 x}_{\mathcal{V}_0} \int
  \frac{\ex^{3\beta_0}}{N} \left( \dot{\beta}_+^2 +\dot{\beta}_-^2
  + 2 \dot{\beta}_0^2 + 
  \ddot{\beta}_0 - 
  \frac{\dot{N}}{N}\dot{\beta}_0
  \right)\dd\tau,
  \label{SEH}
\end{equation}
in which we assume the comoving volume of 3-space to be finite
(compact space ensuring the extrinsic curvature surface term to be
absent) and set to $\mathcal{V}_0$. Noting that 
$$
\frac{\ex^{3\beta_0}}{N} \left( \ddot{\beta}_0
+ 2 \dot{\beta}_0^2 - \frac{\dot{N}}{N}\dot{\beta}_0
  \right) = \frac{\dd}{\dd \tau}
  \left( \frac{\ex^{3\beta_0}}{N} \dot{\beta}_0 \right)
  - \frac{\ex^{3\beta_0}}{N} \dot{\beta}_0^2,
$$
one integrates \eqref{SEH} by each part and discards the boundary
term to obtain the reduced action
\begin{equation}
\mathcal{S}_\textsc{eh}=\frac{3\mathcal{V}_0}{8\pi\GN}
\int\frac{\ex^{3\beta_0}}{N} \left( -\dot{\beta}_0^2 
+\dot{\beta}_+^2 +\dot{\beta}_-^2 \right) \dd\tau = \int
L\left(\beta_i, \dot{\beta}_i\right)\dd\tau ,
\label{Sreduce}
\end{equation}
from which the momenta are found to be
\begin{equation}
p_0 = \frac{\partial L}{\partial \dot{\beta}_0} =
-\frac{3\mathcal{V}_0}{4\pi\GN}\frac{\ex^{3\beta_0}}{N}
\dot{\beta}_0 \qquad \hbox{and} \qquad p_\pm = \frac{\partial
L}{\partial \dot{\beta}_\pm} =
\frac{3\mathcal{V}_0}{4\pi\GN}\frac{\ex^{3\beta_0}}{N}
\dot{\beta}_\pm, \label{Popm}
\end{equation}
leading to the Hamiltonian
\begin{equation}
\mathcal{S}_\textsc{eh} = \int \Bigl[ p_0 \dot{\beta}_0 +
p_+ \dot{\beta}_+ + p_- \dot{\beta}_- -
\underbrace{\frac{2\pi\GN}{3\mathcal{V}_0} \ex^{-3\beta_0}
\left( -p_0^2 + p_+^2 + p_-^2 \right) N}_{H = C N} \Bigr],
\label{HBI}
\end{equation}
where we emphasise the constraint $C$, which classically
vanishes,  with the lapse function $N(\tau)$ always being 
nonvanishing.

For later convenience, we consider instead of $\beta_0$ the
volume variable $V = \exp (3\beta_0)$, with momentum
$p_V=p_0 \exp(-3\beta_0)/3$, transforming the Hamiltonian
into
\begin{equation}
H = \frac{3V}{8} \left( -p_V^2 +
\frac{p_+^2+p_-^2}{9V^2}\right) N = C N. \label{C0}
\end{equation}
In \eqref{C0} and in what follows, we assume units such that
$16\pi\GN = \mathcal{V}_0$.

As $H$ in \eqref{C0} depends on neither $\beta_\pm$, these
cyclic coordinates have conserved associated momenta
$p_\pm$, which we write as
$$
p_+ = k \cos \varphi \qquad \hbox{and} \qquad p_+ = k \sin
\varphi,
$$
in which we assume $k>0$. Correspondingly, we find that the
corresponding momenta can be written as $p_k = -\beta_+
\cos\varphi - \beta_- \sin\varphi$ and $p_\varphi = k
(\beta_+ \sin\varphi - \beta_- \cos\varphi)$. Plugging these
relations into the Hamiltonian, it turns out that the new
variable $\varphi$ can be altogether ignored as neither
$\varphi$ nor its momentum $p_\varphi$ appears in $H$. We
thus end with
\begin{equation}
\mathcal{S}_\textsc{eh} = \int \left[ p_V \dot{V} + p_k
\dot{k} - \frac{3V}{8} \left(-p^2_V + \frac{k^2}{9V^2}
\right) N\right] \dd\tau. \label{Hfin}
\end{equation}
As the volume is positive definite, solving the constraint
$C=0$ translates into setting $k^2 = 9 V^2 p_V^2$, so that
$\dd k/\dd\tau = [1/(2k)] [\dd (k^2)/\dd\tau] = [1/(2k)]
[\dd (9 V^2 p_V^2)/\dd\tau]$, and finally
$$
p_V \dot{V} + p_k \dot{k} = \frac{\dd}{\dd\tau} \left[ V p_V
\ln V + \frac12 V^2 p_V^2 \left( \frac{9 p_k}{k} - \frac{\ln
V}{Vp_V} \right) \right] - \frac12 V^2 p_V^2
\frac{\dd\Upsilon}{\dd\tau},
$$
where
\begin{equation}
\Upsilon = \frac{9 p_k}{k} - \frac{\ln V}{Vp_V}.
\label{UpsDef}
\end{equation}
The variable $\Upsilon$
now serves as an integrating measure in the action, and it has
therefore turned into a clock variable.

As now the constraint is satisfied; setting aside the
boundary term above, one finally obtains the action in the
form
\begin{equation}
\mathcal{S}_\textsc{eh} = -\frac12 \int  V^2 p_V^2
\dd\Upsilon = \int \left[ V p_V \dd (V p_V T) -\frac12 V^2 p_V^2
\dd (\Upsilon + T) -\frac12 \dd (V^2 p_V^2 T)\right],
\label{T1}
\end{equation}
where we have introduced an arbitrary function
$T(V,p_V,p_k)$ of the original relevant variables.
Discarding the last, integrated term and setting $q = V p_V
T$ and $p=V p_V$, the action is expressed in the canonical
form
\begin{equation}
\mathcal{S}_\textsc{eh} = \int \left[ p \frac{\dd q}{\dd t}
-H(q,p) \right] \dd t= \int \left( p \frac{\dd q}{\dd t}
-\frac12 p^2 \right) \dd t,
\label{T2}
\end{equation}
provided we set $t=\Upsilon + T$ as the new time variable.

A thorough discussion of this issue together with that of
choosing the otherwise arbitrary function $T$ is given in
Ref.~\cite{Malkiewicz:2019azw}, where in particular it was
shown that there exist two categories of possible choices,
namely the so-called fast- and slow-time gauges. In the
former case, the singularity is somehow not removed upon
quantisation, in the sense that the wavefunction
asymptotically shrinks towards a $\delta-$function around the 
vanishing scale factor (hence a singularity) after an
infinite amount of time. In the latter case of slow-time
gauge, the singularity is resolved into a bouncing universe.

We shall restrict out attention in what follows to the slow-time
gauge only and therefore assume the arbitrary function to
take the simple form $T=V^{-1}$, leading the relevant
variable $q$ to be identified with the volume $V$. The
classical Hamiltonian is now reduced to that of a free
particle confined to the semi-infinite half line $\mathbb{R}^+$.
We now turn to the quantisation of this problem.

\section{Affine Quantisation} \label{sec_affine}

Quantising a Hamiltonian system in principle follows a
well-defined procedure, referred to as ``canonical
quantization'' and proposed by Dirac. It consists of 
replacing the relevant dynamical variables by corresponding
operators and the Poisson brackets by $i$ times the
commutators between these operators. In the position
representation with wavefunction $\Psi(q,t)$, the operator
$\hat{Q}$ becomes the multiplication by $q$ and the momentum
yields $\hat{P}\Psi =-i\hbar \partial \Psi /\partial q$.

Canonical quantisation is based on the unitary and
irreducible representation of the group of translations in
the $(q,p)$ plane, the Weyl--Heisenberg group. For a particle
living in a smaller space, it therefore might not apply in a
straightforward manner, as one has to reduce the Hilbert
space of available states to ensure the mathematical
properties of the observables to be satisfied. Instead of
adopting this potentially problematic approach, we propose
that the so-called covariant integral be considered. This is
based on a minimal group of canonical transformations with
a nontrivial unitary representation.

For the half-plane that arises in the Bianchi I case of the
previous section, the natural choice is the 2-parameter
affine group of a real line with elements
$(q,p)\in\mathbb{R}^+\times\mathbb{R}$, transforming
$s\in\mathbb{R}$ into $(q,p)\cdot s = s/q + p$ and with
composition law
\begin{equation}
\left\{ \left( q_0,p_0\right),\left( q,p\right)\right\}
\mapsto \left( q',p'\right)  = \left( q_0,p_0\right)
\circ \left( q,p\right) = 
\left( q_0 q,\frac{p}{q_0}+p_0\right)
\end{equation}
and left-invariant measure $\dd q' \wedge \dd p' = \dd q
\wedge \dd p$. It is clear that $q$ represents a change of
scale, which is what one would expect for a scale factor
(dimensionless in our conventions),
while the momentum is rescaled and translated as the scale
is modified. For the 2-parameter affine group, one can find
a unitary, irreducible and square-integrable representation
in the Hilbert space $\mathcal{H}=L^2(\mathbb{R}^+,\dd x)$.
It reads
\begin{equation}
\langle x|U(q,p)|\zeta\rangle = \langle x|q,p\rangle_\xi
=\frac{\ex^{ipx/\hbar}}{\sqrt{q}}\xi\left(
\frac{x}{q}\right),
\label{qpx}
\end{equation}
where $\xi(x)=\langle x | \xi\rangle\in \mathcal{H}$ and
$|\xi\rangle \in \Hu$ is an (almost) arbitrary fiducial state vector
belonging to the Hilbert space (see
Ref.~\cite{Martin:2021dbz} and references therein). As for
the unitary operator $U(q,p)$ implementing an affine
transformation, it reads
\begin{equation}
U(q,p)=\ex^{ip\widehat{Q}/\hbar}\ex^{-i\ln q\widehat{D}/\hbar},
\label{Uqt}
\end{equation}
with $\widehat{D} := \frac12 (\widehat{Q}\widehat{P}+
\widehat{P}\widehat{Q})$  as the dilation operator, forming
with $\widehat{Q}$ the algebra $\left [ \widehat{Q},\widehat{D}
\right] = i\hbar \widehat{Q}$.

Let us define the series of integrals
\begin{equation}
\rho_\xi(s) := \int \frac{\langle \xi| x\rangle \langle x
|\xi \rangle}{x^{s+1}}\, \dd x= \int
\frac{|\xi(x)|^2}{x^{s+1}}\, \dd x \, < \infty \quad
\hbox{and} \quad \sigma_\xi(s) :=\int \left| \frac{\dd
\xi(x)}{\dd x} \right|^2 \frac{\dd x}{x^{s+1}}
\label{rhosigma}
\end{equation}
assumed convergent, and the quantisation rule

\begin{equation}
f(q,p)\mapsto A_\xi[f] := \mathcal{N}_\xi
\int_{\mathbb{R}^+\times\mathbb{R}} \frac{\dd q\dd p}{2\pi
\hbar}|q,p\rangle_\xi\, f(q,p)\, \prescript{}{\xi}\langle
q,p| \quad \hbox{with} \quad \mathcal{N}_\xi =
\frac{1}{\rho_\xi(0)} =: \frac{1}{\rho_0}~, \label{quantiz}
\end{equation}
associating to each function $f(q,p)$ of the classical dynamical
variables a unique operator  $A_\xi[f]$ in the Hilbert space $\Hu$.
The normalisation $\mathcal{N}_\xi$ comes from the resolution
of unity
\begin{equation}
\int \frac{\dd q \dd p}{2\pi \hbar \rho_0} | q,p\rangle_\xi
\prescript{}{\xi}{\langle q,p|}
= \int \dd x |x\rangle\langle x| = \mathbbm{1} = A_\xi[1],
\label{Un}
\end{equation}
using $2 \pi\hbar\delta (x-y) = \int \ex^{i p (x-y)/\hbar}\dd p$.
Useful operators can then be represented, such as powers
of $q$ or the momentum $p$, namely
\begin{equation}
A_\xi[q^s] = \int_{\mathbb{R}^+\times\mathbb{R}} \frac{\dd
q\dd p}{2\pi \hbar\rho_0} |q,p\rangle_\xi\, q^s\,
\prescript{}{\xi}{\langle q,p|} = \frac{\rho_\xi(s)}{\rho_0}
\widehat{Q}^s
\label{AQs}
\end{equation}
and
\begin{equation}
A_\xi[p] =
\int_{\mathbb{R}^+\times\mathbb{R}} \frac{\dd q\dd p}{2\pi
\hbar\rho_0} |q,p\rangle_\xi\, p\, \prescript{}{\xi}{\langle
q,p|} q,p| =\widehat{P}, 
\label{Ap}
\end{equation}
showing that the fiducial state $|\xi\rangle$ should be such
that $\rho_\xi(1) = \rho_\xi(0)$ in \eqref{rhosigma} to
ensure the canonical commutation relations
$[A_\xi[q],A_\xi[p]] = [\widehat{Q}, \widehat{P}] = i\hbar$.
Finally, the compound quantity $q^s p^2$ is quantised to
\begin{equation}
A_\xi[q^s p^2] = \frac{\rho_\xi(s)}{\rho_0} 
\widehat{P}\widehat{Q}^s\widehat{P} + \hbar^2\left[
\frac{s (1-s) \rho_\xi(s)}{2\rho_0} +
\frac{\sigma_\xi(s-2)}{\rho_0}\right]
\widehat{Q}^{s-2},
\label{qsp2}
\end{equation}
so that the classical Hamiltonian in \eqref{T2}, namely
$H(q,p) = \frac12 p^2$, has an affine quantum counterpart
given by
\begin{equation}
A_\xi[H(q,p)] = \widehat{H}(\widehat{Q},\widehat{P}) =
\frac12 \widehat{P}^2 +\hbar^2 \Ka_\xi \widehat{Q}^{-2}
= \frac12 \widehat{P}^2 +V(\widehat{Q}),
\label{Hquantum}
\end{equation}
with $\Ka_\xi =\sigma_\xi(-2)/\rho_0$; given the
arbitrariness of the fiducial vector, this coefficient is
essentially arbitrary. If instead of the affine quantisation
one applies the canonical prescription, it would simply
vanish ($\Ka_\mathrm{can} \to 0$). Among the advantages of
this quantisation is the fact that it permits us  to merely
parametrise the well-known operator ordering ambiguity,
replacing it by a single unknown number, to be ultimately
fixed by experiment.

It should be noted that if $\Ka_\xi \geq \frac34$, the
Hamiltonian \eqref{Hquantum} is essentially self-adjoint, so
one needs not impose any boundary conditions at $q=0$, the
dynamics generated being unique and unitary by
construction~\cite{Vilenkin:1987kf}. In the framework of
quantum cosmology that concerns us here, affine
quantisation induces a repulsive potential $V(\widehat{Q})$
thanks to which it is natural to expect that the classical
GR Big Bang singularity will be resolved by quantum effects,
as indeed is found to happen with our choice of
clock~\cite{Malkiewicz:2019azw}.

\section{Clock Transformations} \label{sec_clock}

\textls[-15]{A classically constrained Hamiltonian theory with
$H_\text{full}(q_\text{full},p_\text{full}) \approx 0$ and
deparametrised} to a reduced phase space $(q,p)$ using an
internal degree of freedom $t$ as clock is invariant under
the so-called clock transformations. The idea behind the
clock transformation is the following: given a clock $t$ and
its associated reduced phase space formalism $(q,p,t)$, one
seeks, prior to deparametrisation, another choice of clock
$\tilde{t}$, say, leading to a similar reduced phase space
formalism $(\tilde{q},\tilde{p},\tilde{t})$. This involves
transformations of both the clock variable $t\mapsto
\tilde{t}(q,p,t)$ and the canonical variables $(q,p)\mapsto
[\tilde{q}(q,p,t), \tilde{p}(q,p,t)]$ as the change in time
generally changes the canonical relations in reduced phase space.
These clock transformations can also be  understood as
canonical transformations in the full phase space
$(q_\text{full}, p_\text{full})$, thereafter restricted to
the constraint surface. This restriction is responsible for
altering the canonical relations in the reduced phase space.
The relation between the full- and reduced-phase-space
formulations of the clock transformations was investigated
in~\cite{Malkiewicz:2014fja}.

Let us start by noticing that the new canonical variables
$(\tilde{q},\tilde{p})$ associated with the new clock
$\tilde{t}$ can be chosen conveniently as to satisfy
\begin{align}
\ud q\wedge\ud p -\ud t \wedge \ud H(q,p,t) = 
\ud \tilde{q} \wedge
\ud \tilde{p} - \ud \tilde{t} \wedge \ud
H(\tilde{q},\tilde{p},\tilde{t}),
\end{align}
where the form of the reduced Hamiltonian
$H(\cdot,\cdot,\cdot)$ is preserved by the clock
transformation~\cite{Malkiewicz:2017cuw}. The above choice
of $\tilde{q}$ and $\tilde{p}$ is convenient because once
the solution to the dynamics is known in $t$ as
$q=S_q(t),p=S_p(t)$, it is automatically known in all other
clocks $\tilde{t}$ as
$\tilde{q}=S_q(\tilde{t}),~\tilde{p}=S_p(\tilde{t})$. It is
obviously the same physical solution but now differently
parametrised. It is easy to notice that $ \tilde{t}\neq t$
implies  $\ud q\wedge\ud p\neq \ud \tilde{q}\wedge\ud
\tilde{p}$, and the clock transformation indeed alters the
canonical relations in the reduced phase space. This is a
sufficient reason for the existence of unitarily
inequivalent quantum dynamics based on different choices of
clock (as we shall see shortly). Let
us now explain how the clock transformation satisfying the
above condition is determined in practice.

One first calculates Dirac observables $C_i(q,p,t)$ ($i=1,2$
in the two-dimensional phase space discussed here) by solving
their defining equation, namely $\partial_t C_i + \{C_i,  H
\}_\textsc{p.b.} =0$. One then demands that for the
transformation $(q,p,t) \mapsto
(\tilde{q},\tilde{p},\tilde{t})$, one has $C_i(q,p,t) = C_i
(\tilde{q},\tilde{p},\tilde{t})$, thus leading to the
required relationship between $(\tilde{q},\tilde{p})$ and
$(q,p)$ for an arbitrary change of clock time $t\mapsto
\tilde{t}$. It should be emphasised at this point that the
clock transformation provides an actual invariance provided
there is an underlying Hamiltonian, even a time-dependent
one.

Consider first the Hamiltonian in \eqref{T2}, namely
$H_0=\frac12 p^2$. The Dirac observable requirement then
reads $\partial_t C_i + p \partial C_i/\partial V = 0$. One
set of solution is $C_1=p$ and $C_2=pt-q$, leading to
$\tilde{p} = p$, and $\tilde{p}\tilde{t}-\tilde{q}=pt-q$,
which implies $\tilde{q} = q + (\tilde{t} - t ) p = q+\Delta
p$, thereby defining the function
\begin{equation}
\Delta(t,q,p):=\tilde{t}-t.
\label{Delta}
\end{equation}
The effects of this transformation was studied in
Ref.~\cite{Malkiewicz:2019azw} for various arbitrary
$\Delta$. In phase space, the solutions for $H_0$ are
$\dot{p} = 0$, and therefore $p=p_0$ constant, with
$\dot{q}= p = p_0$ so that $q=p_0 t + q_0$: these are
straight lines in the $(q,p)$ space, labelled by $t$.
Changing to $\tilde{t}$ yields the same Hamiltonian, now in
the new variables, and therefore the same equations of motion, and thus
the same formal solutions, namely $\tilde{p} =\tilde{p}_0$
and $\tilde{q}=\tilde{p}_0 \tilde{t} + \tilde{q}_0$.
Applying the transformation implies $\tilde{q}_0=q_0$ and
$\tilde{p}_0=p_0$. For one particular solution,
Equation~\eqref{Delta} provides $\tilde{t} (t)$, which must be
monotonic and invertible, yielding $t(\tilde{t})$: $q(t)$
now transforms into $q(\tilde{t})$, and because $p$ is
constant, one recovers straight lines, now labelled in a
different way.

Let us now turn to the more complicated example of the
quantum Hamiltonian \eqref{Hquantum}, now considered
classical and written as $H=\frac12 p^2 + \mathcal{K}/q^2$.
One now needs to find the solution to
$$
\frac{\partial C_i}{\partial t} + \frac{2\Ka}{q^3} 
\frac{\partial C_i}{\partial p} + p \frac{\partial C_i}{\partial q} =0,
$$
which is solved by the set
\begin{equation}
C_1 = \frac12 p^2 + \frac{K}{q^2} = H(q,p) \quad \hbox{and}
\quad C_2 = qp-2H(q,p) t.
\label{C1C2}
\end{equation}
For the clock transformation, one derives from \eqref{C1C2}
the relations
\begin{equation}
\tilde{q}^2 = q^2 + Z \quad \hbox{and} \quad 
\tilde{p}^2 = p^2 +\frac{2\Ka Z}{q^2 (q^2 + Z)} = 2 \left(
H-\frac{\Ka}{q^2+Z}\right),
\label{QtPt}
\end{equation}
where we have set $Z = 2\Delta (pq+H\Delta)$. Two conditions
must be imposed for the choice of
$\Delta$. First, it must be made such that it satisfies $Z\geq -q^2$ to
ensure $\tilde{q}^2\geq 0$ and hence 
$\tilde{q}\in \mathbb{R}$. Second, the inequality
$$
1+\frac{\partial \Delta}{\partial t} +
\{\Delta,H\}_\textsc{p.b.} =
1+\frac{\partial \Delta}{\partial t} +
p\frac{\partial\Delta}{\partial q}
+\frac{2\Ka}{q^3} \frac{\partial\Delta}{\partial
p} \not= 0
$$
must hold in order that the time delay function $\Delta$
ensures monotony of the new time with respect to the old
one, i.e., $\dd\tilde{t}/\dd t >0$. Note that the
transformation \eqref{QtPt} gives back that corresponding to
$\Ka=0$ in both limits $\Ka\to0$ and $q\to\infty$.

Our system originates classically from the simplest option,
namely $H\to H_0 =\frac12 p^2$, but that derived from the
quantum one \eqref{Hquantum} can imply semiclassical (or
perhaps semiquantum~\cite{Martin:2021dbz}) trajectories that
should be invariant under \eqref{QtPt}. It is therefore
important to derive actual trajectories one way or another
to be able to estimate the effects a choice of clock \mbox{can
have}.

\section{Trajectories} \label{sec_traj}

\begin{figure}[t]

\centering
\includegraphics[width=6.85cm]{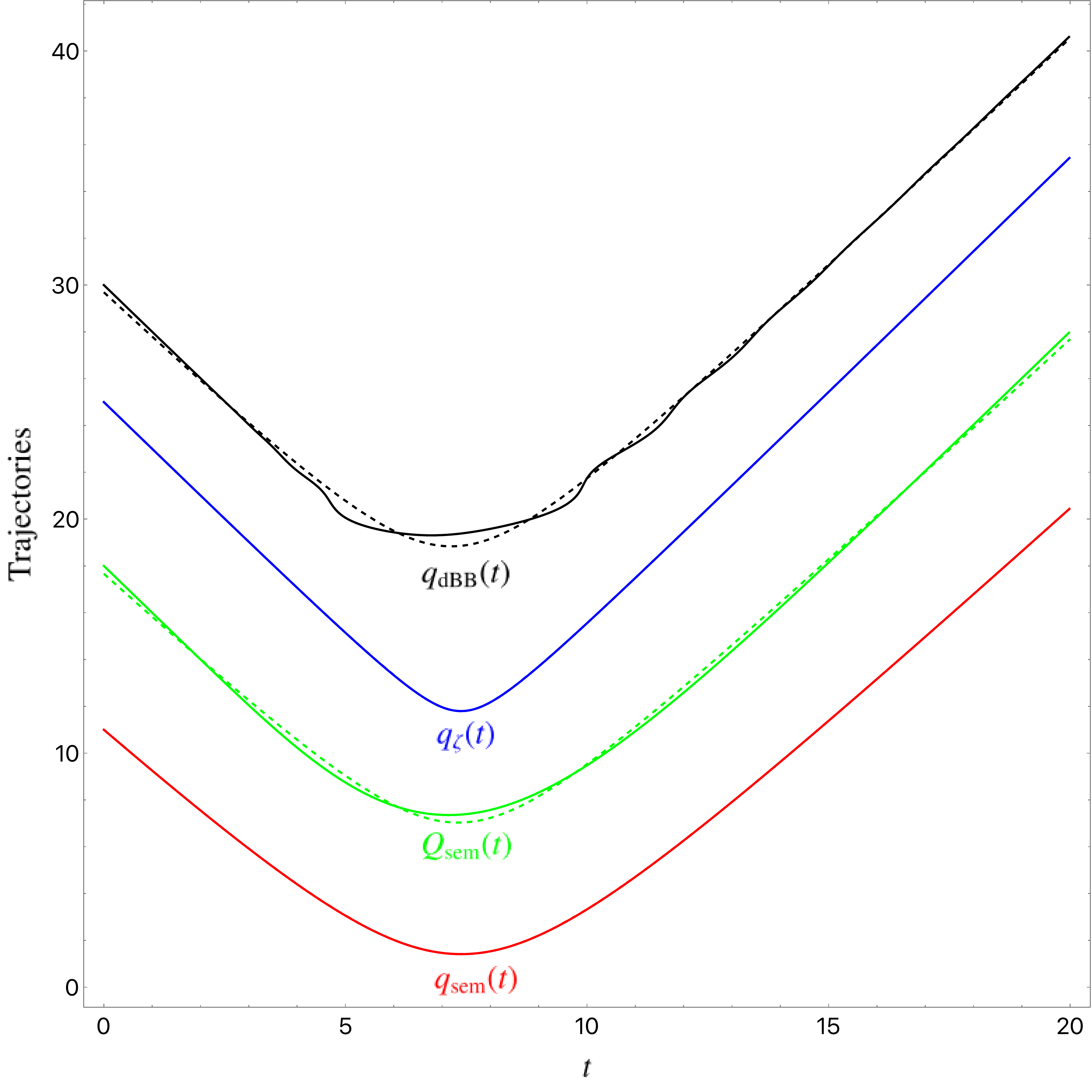}
\includegraphics[width=6.85cm]{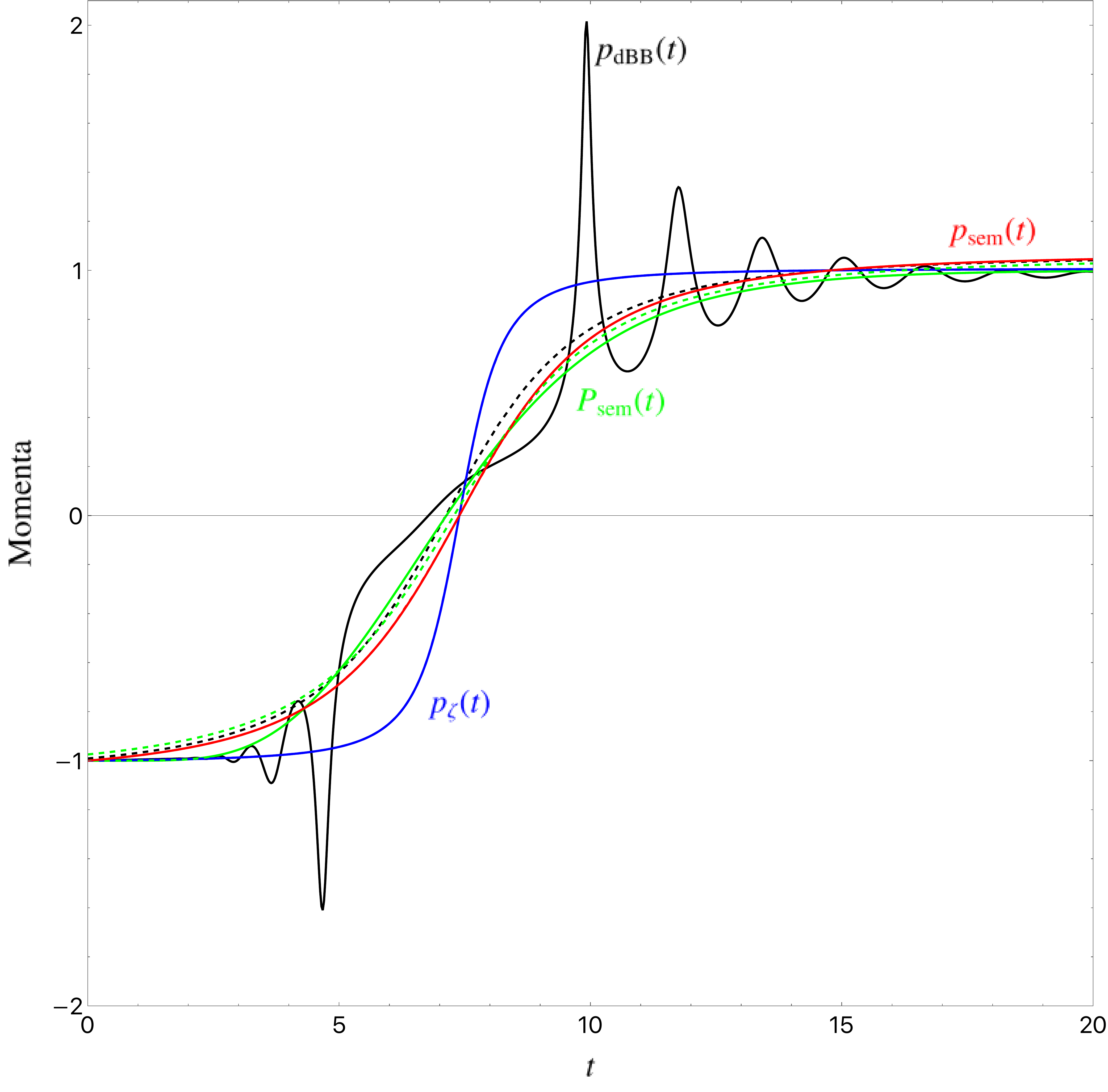}

\caption{Time developments of the various trajectories
proposed in the text (left panel) as obtained from the
wavefunctions of Figure~\ref{psis}. Except for the dBB case,
all the definitions used for semiclassical trajectories are
well fitted (or exactly given) by the solution
\eqref{solsem}, shown as dashed lines for each curve (these
have been arbitrarily displaced up and down for visual
purposes; otherwise they are hardly distinguishable). The
right panel shows the relevant associated momenta and
emphasises the large discrepancy visible only in the dBB
case.}

\label{trajs}
\end{figure}

There are various ways to implement physically meaningful
trajectories in our quantum description of the dynamics of a
Bianchi I universe, as illustrated in Figure~\ref{trajs}.
The first and most obvious consists
merely in evaluating expectation values. If the wavefunction
is sufficiently narrow, this can provide an effective
semiclassical approximation.

With the Hamiltonian \eqref{Hquantum}, it has been shown
that an approximate  space trajectory can be deduced
directly from the quantum version of the
algebra~\cite{Malkiewicz:2019azw}: using
$[\widehat{D},\widehat{H}] = 2 i \widehat{H}$ and the fact
that $\widehat{H}$ is a constant operator, one can integrate
the Heisenberg equation of motion $\dd\widehat{D}/\dd t = -i
[\widehat{D},\widehat{H}] = 2 \widehat{H}$, leading to
$\widehat{D}(t) = \widehat{D}(0) + 2 \widehat{H} t $. Even
though the operator $\widehat{Q}$ itself cannot be
integrated directly from the algebra because
$[\widehat{Q},\widehat{H}] = i \widehat{P}$ and
$[\widehat{P},\widehat{H}] = 2 i \Ka\widehat{Q}^{-3}$, its
square leads to $[\widehat{Q}^2,\widehat{H}] = 2 i
\widehat{D}$, so one finds $\dd\widehat{Q}^2/\dd t = -i
[\widehat{Q}^2,\widehat{H}] = 2 \widehat{D}(t) =
2\widehat{D}(0) + 4 \widehat{H} t $. This implies
$\widehat{Q}^2 = \widehat{Q}^2(0) + 2 [ \widehat{D}(0) t +
\widehat{H} t^2]$. A semiclassical trajectory can then be
defined in phase space by setting $q_\mathrm{sem} (t) =
\sqrt{\langle \widehat{Q}^2 \rangle}$ and $p_\mathrm{sem}
(t) = \langle \widehat{D} \rangle/q_\mathrm{sem} (t)$.
Shifting the time to set the minimum of $q_\mathrm{sem} (t)$
at $t_\textsc{b} = 0$, one obtains a bouncing behaviour 
$q_\mathrm{sem} (t) = q_\textsc{b} \sqrt{(\omega t)^2+1}$.

\begin{figure}[t]
\centering
\includegraphics[width=4.5cm]{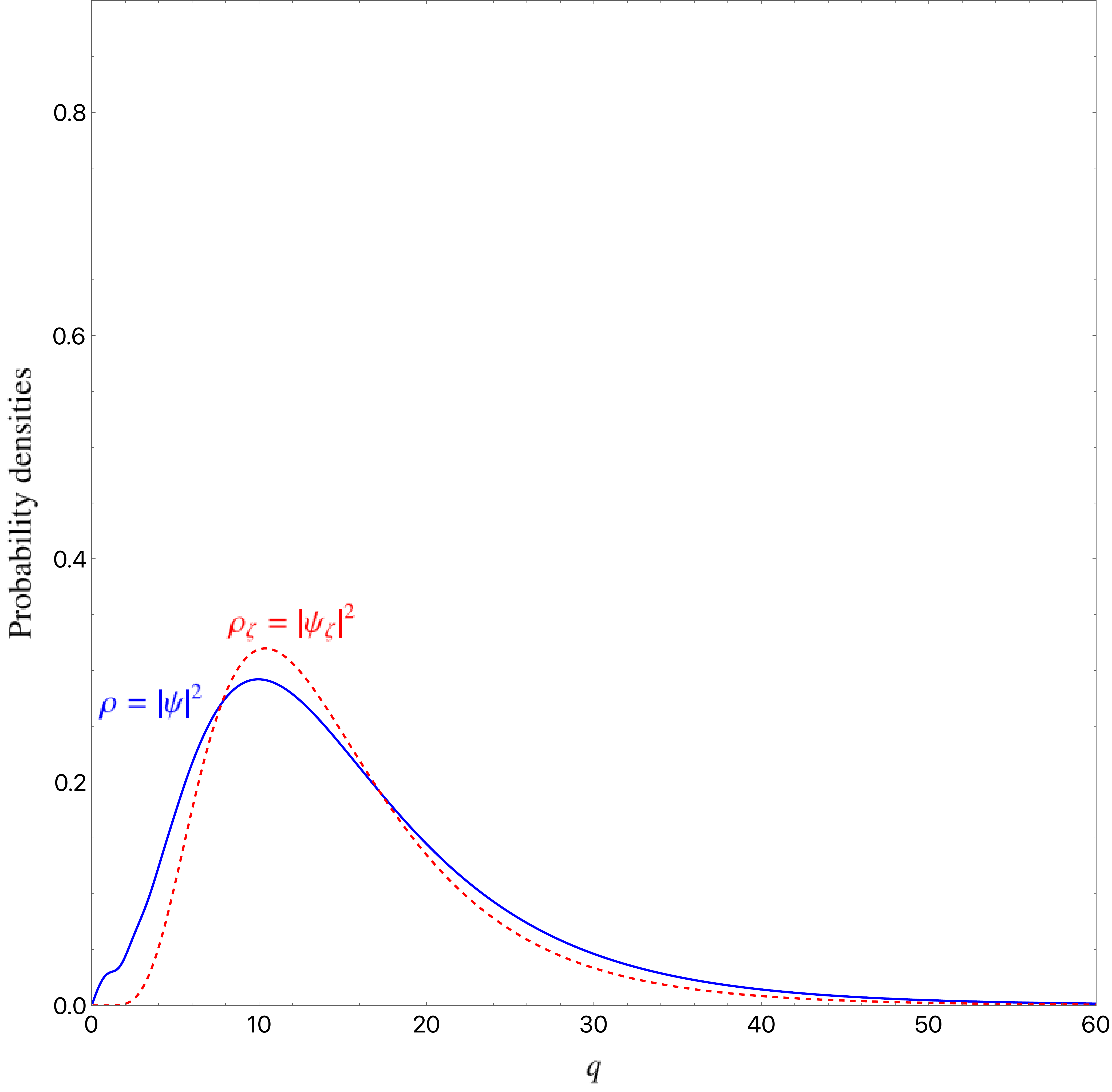}
\includegraphics[width=4.5cm]{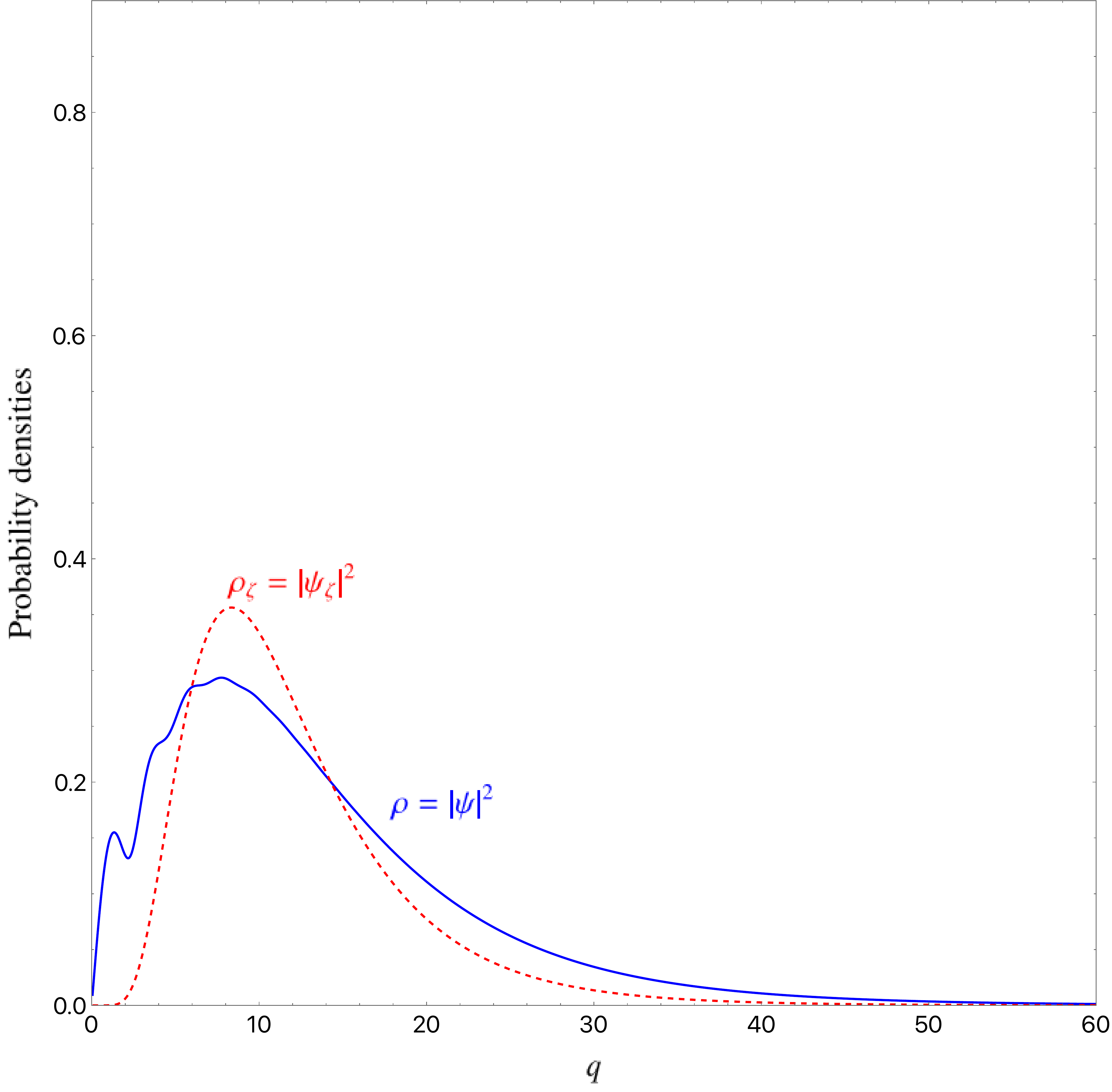}
\includegraphics[width=4.5cm]{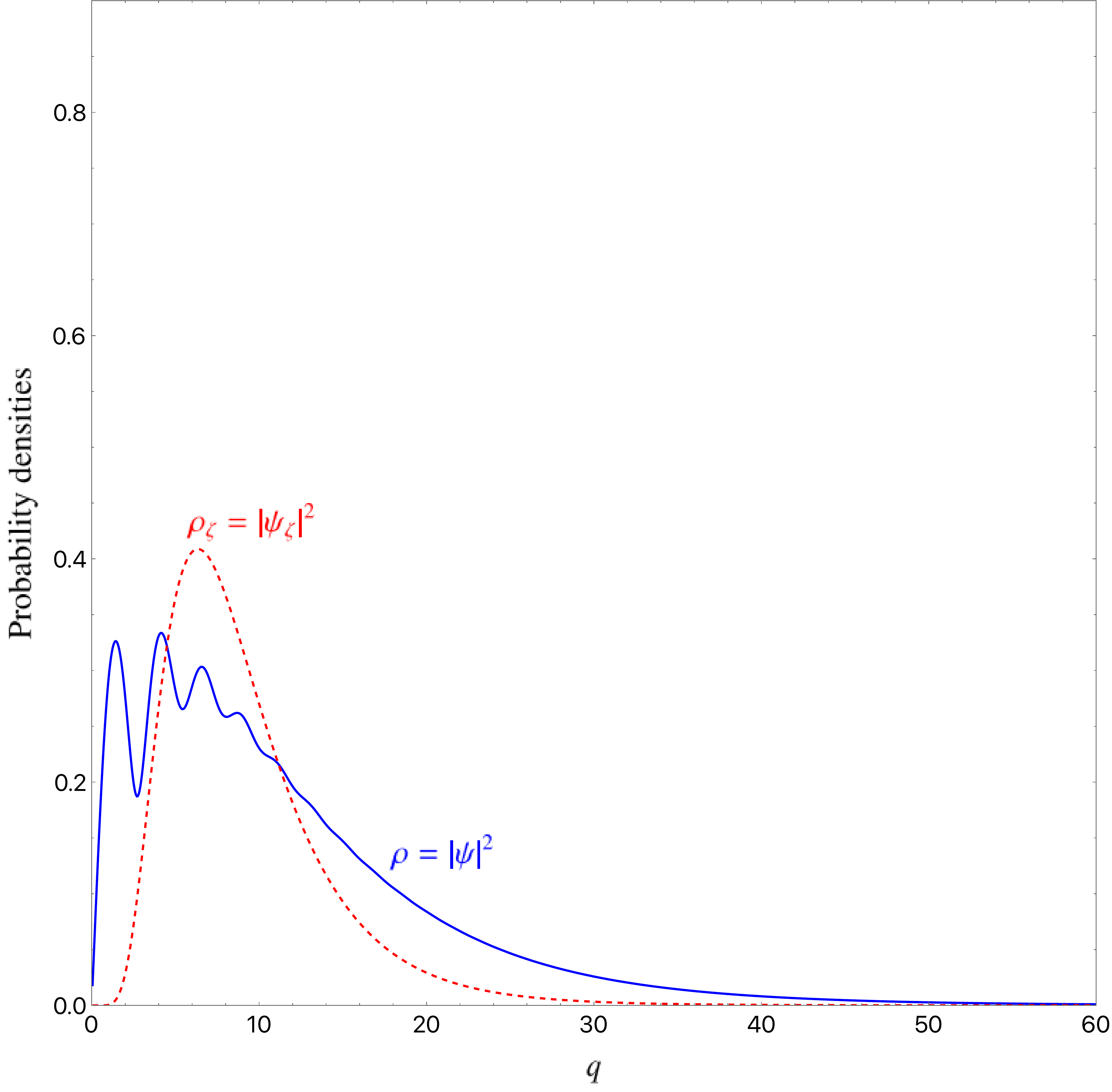}
\includegraphics[width=4.5cm]{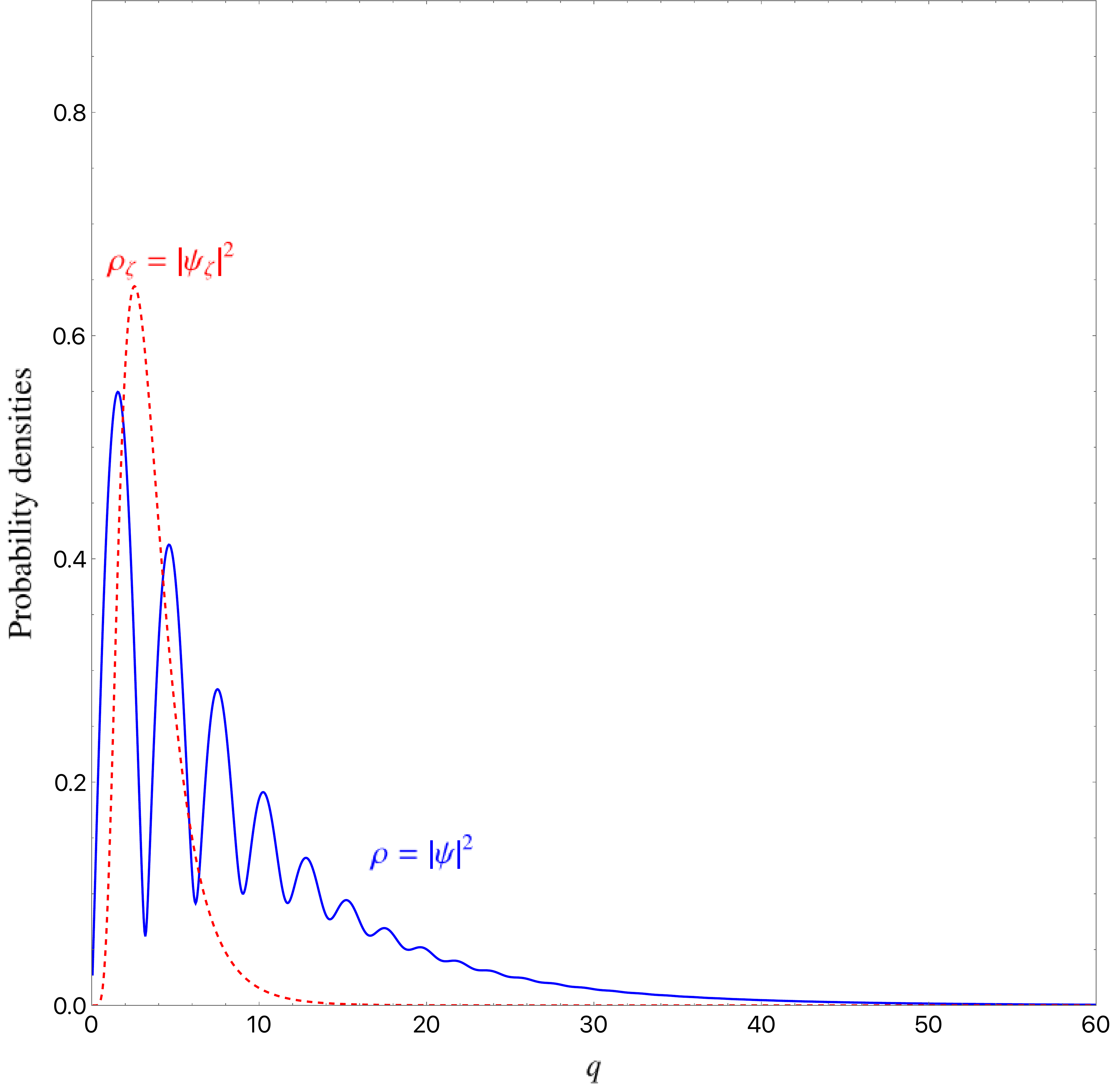}
\includegraphics[width=4.5cm]{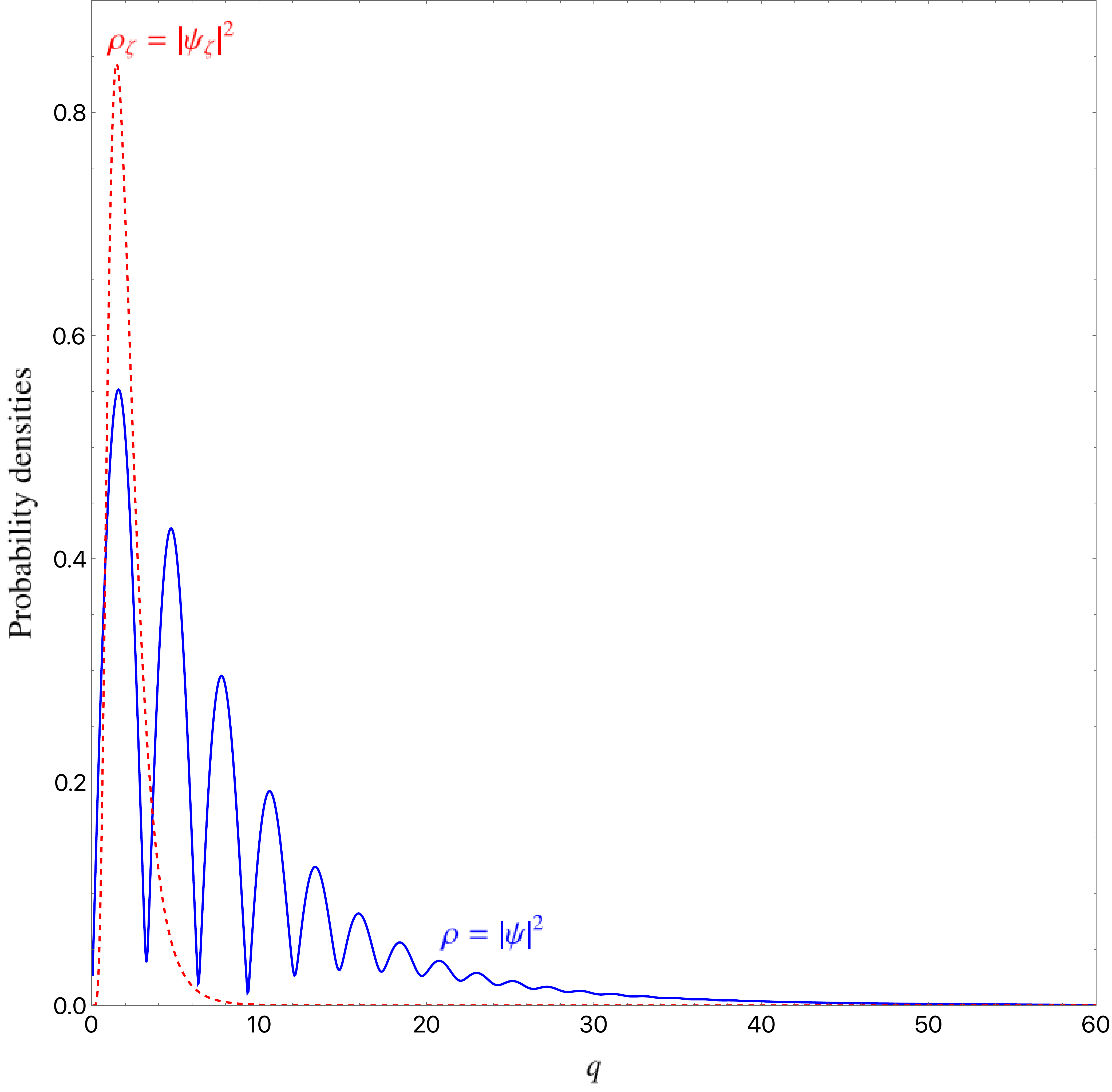}
\includegraphics[width=4.5cm]{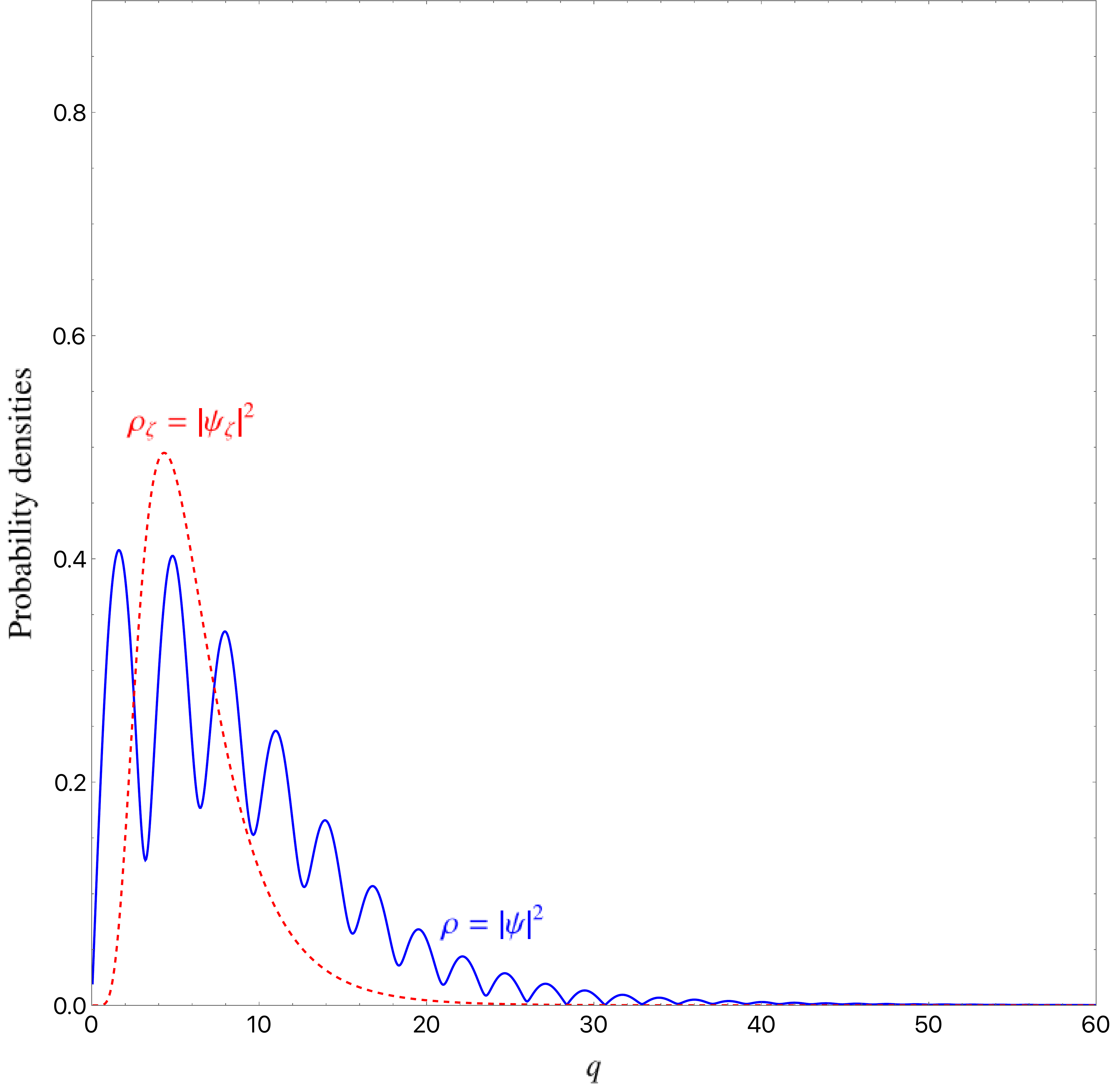}
\includegraphics[width=4.5cm]{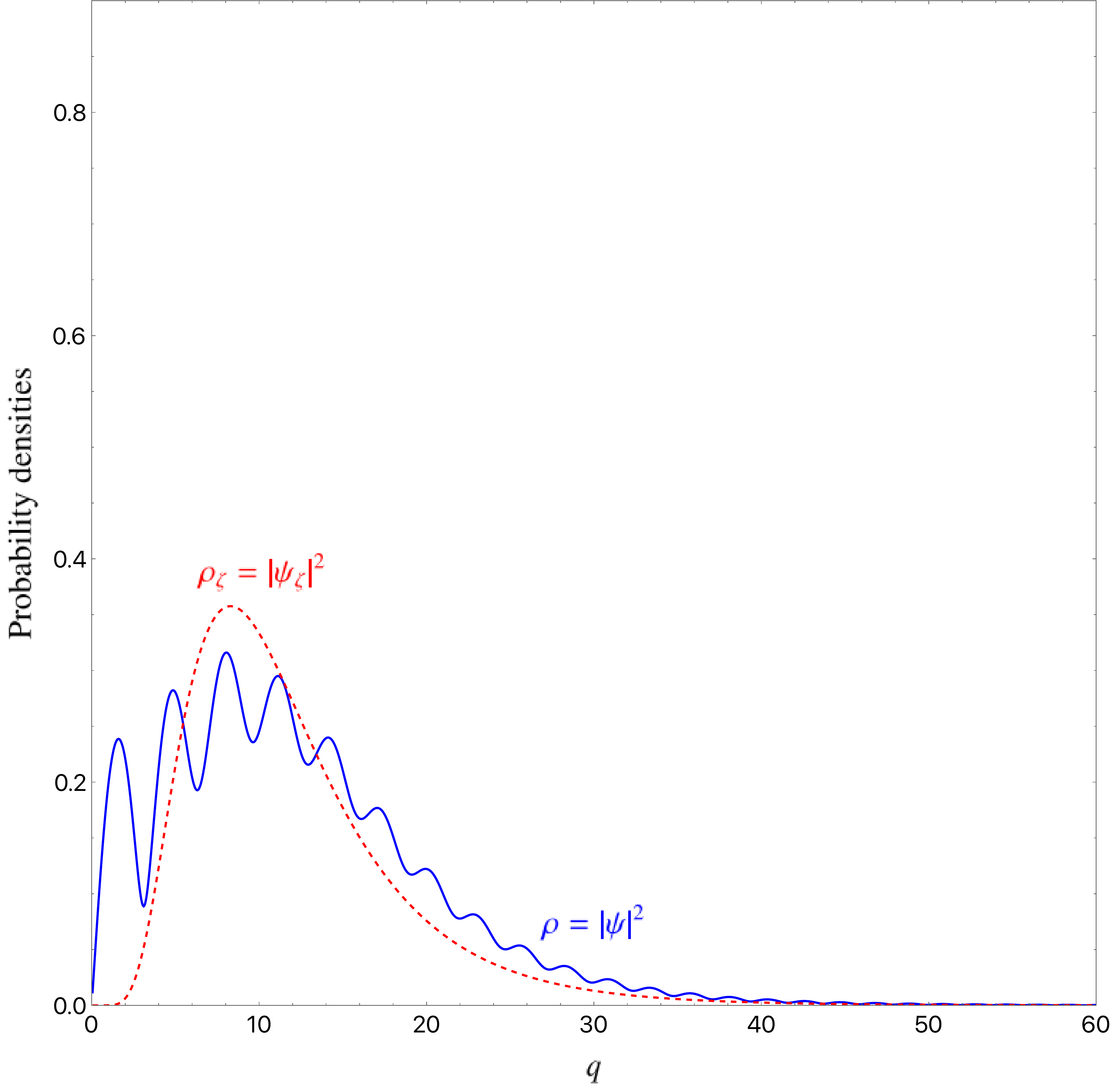}
\includegraphics[width=4.5cm]{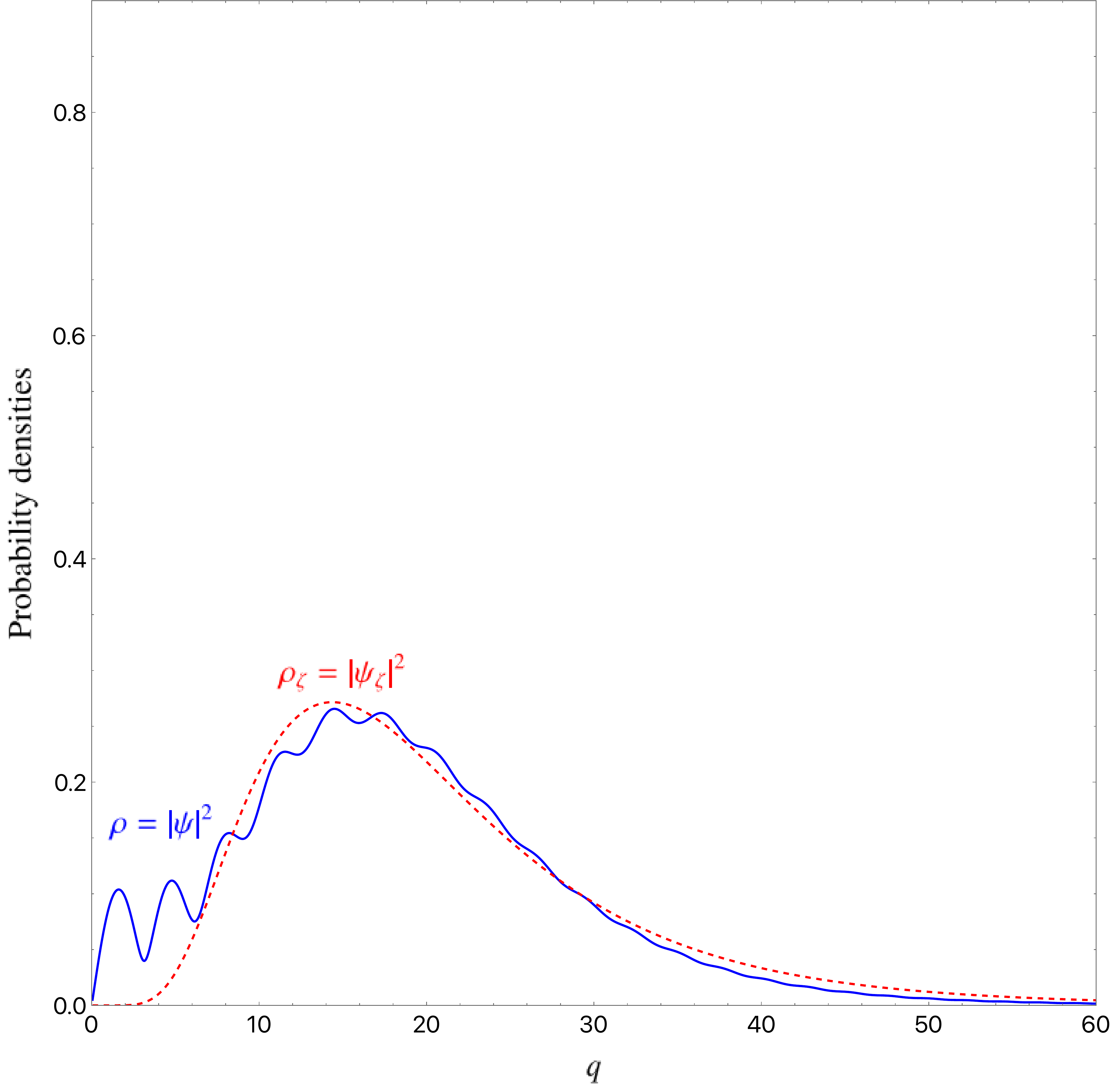}
\includegraphics[width=4.5cm]{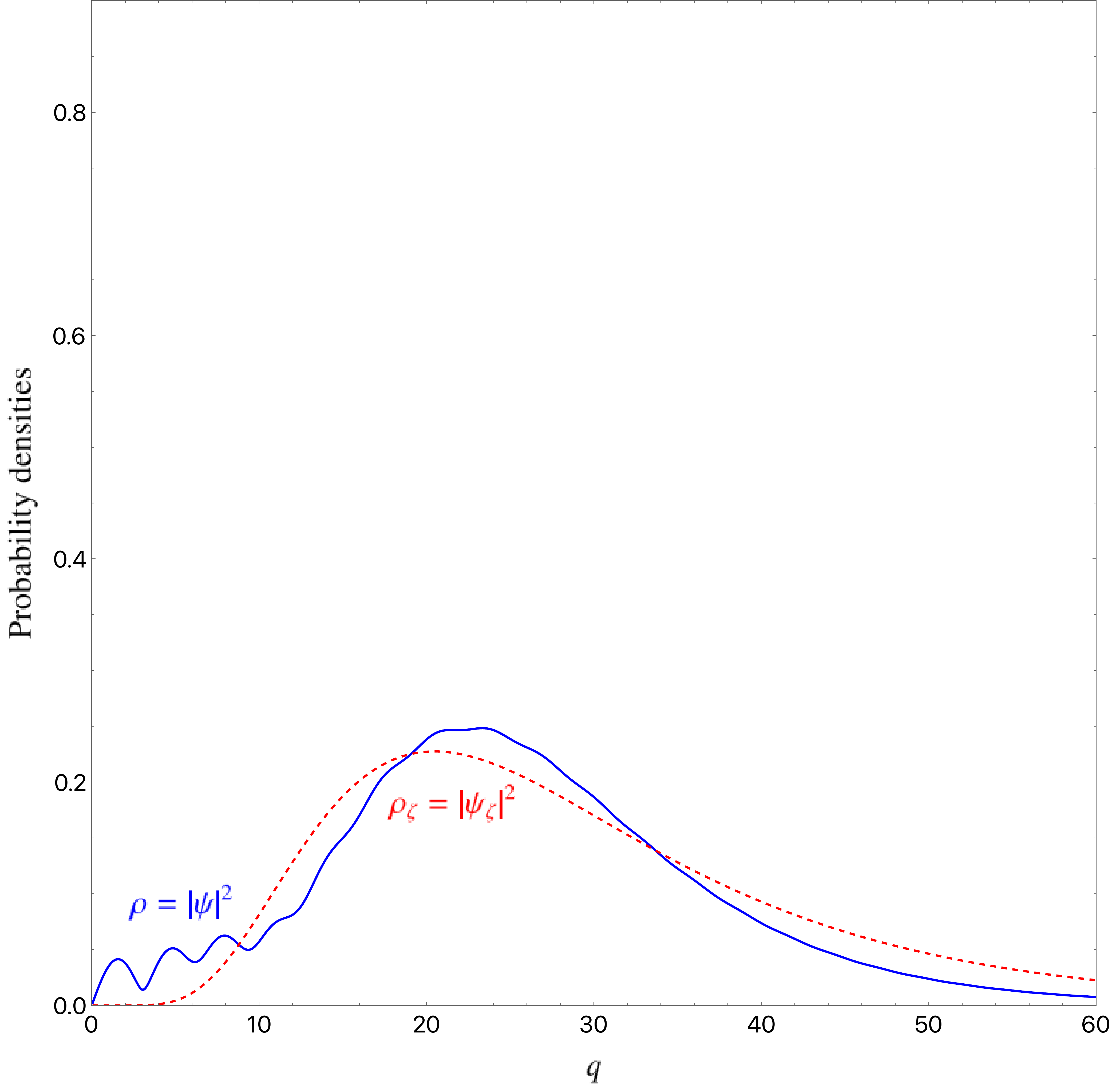}

\caption{Snapshots of the time evolution of both
the full wavefunction $\psi(q,t)$ and its
coherent state approximation $\psi_\zeta(q,t)$,
with the initial condition shown in Figure~\ref{phase}.
As the wavefunctions approach the origin $q\to0$,
the true solution $\psi$ starts oscillating, with the
oscillations developing further with time until
the wave packet is far enough from the origin and
they begin to be damped. On the other hand, the
coherent state wavefunction $\psi_\zeta$ never
oscillates, being merely squeezed at the origin
and then bouncing away. In the large time limit,
they both evolve in more and more similar ways so
that $\lim_{t\to\infty} \psi \sim \psi_\zeta$.}

\label{psis}
\end{figure}

Another option consists in solving the Schr\"odinger
equation and evaluating the expectation values directly with
the relevant wavefunction. This leads to another
semiclassical trajectory $Q_\mathrm{sem} (t) = \langle
\widehat{Q}\rangle$ and $P_\mathrm{sem} (t) = \langle
\widehat{P}\rangle$. It turns out that for $t<0$, one has
$Q_\mathrm{sem} (t) \simeq q_\mathrm{sem} (t) $, although
close to the bounce and afterwards, there is a systematic
shift between $Q_\mathrm{sem} (t)$ and $q_\mathrm{sem} (t)
$. The phase space trajectories
$(q_\mathrm{sem},p_\mathrm{sem})$ and
$(Q_\mathrm{sem},P_\mathrm{sem})$ are in good agreement,
with only a difference in their time labelling.

A third way to obtain approximate trajectories consists in
considering coherent states, as defined through
Equations~\eqref{qpx} and \eqref{Uqt}. Indeed, if one changes the
fiducial state $|\xi\rangle$, satisfying the canonical
condition $\rho_\xi(1)=\rho_\xi(0)$ below \eqref{Ap}, to
$|\zeta\rangle$ such that 
$\langle\zeta|\widehat{Q}|\zeta\rangle =1$ and
$\langle\zeta|\widehat{P}|\zeta\rangle =0$,
the Schr\"odinger action
\begin{equation}
\mathcal{S}_\mathrm{sch} [|\psi\rangle] = \int \langle\psi |
\left( i\hbar \frac{\partial}{\partial t} - H \right) |\psi
\rangle \dd t\label{SchAct}
\end{equation}
is transformed into~\cite{Klauder:2015ifa}

\begin{equation}
\begin{aligned}
\mathcal{S}_\mathrm{sch} [|q(t),p(t)\rangle_\zeta] & = \int
\left[ p_\zeta \dot{q}_\zeta -
\prescript{}{\zeta}{\langle q(t),p(t)|} H |q(t),p(t)\rangle_\zeta
\right] \dd t \\ & \to \int \left[ p_\zeta(t) \dot{q}(t) -
H_\mathrm{sem} (q_\zeta(t),q_\zeta(t))\right] \dd t
\end{aligned}
\label{SchHamil}
\end{equation}
once the arbitrary state $|\psi\rangle$ is replaced by the
coherent state $|q(t),p(t)\rangle_\zeta$, now defined with a
priori unknown functions of time $q_\zeta(t)$ and
$p_\zeta(t)$. It is clear from Equation~\eqref{SchHamil} that the
initially arbitrary functions $q_\zeta(t)$ and $p_\zeta(t)$ are now, in
order to minimise the action, subject to Hamilton equations
\begin{equation}
  \dot{q}_\zeta=\frac{\partial {H}_\text{sem}}{\partial
  p_\zeta} \quad \text{and} \quad
  \dot{p}_\zeta=-\frac{\partial {H}_\text{sem}}{\partial
  q_\zeta}. \label{eomsem}
\end{equation}
with the original Hamiltonian replaced by the semiclassical
one $H_\mathrm{sem}$.

\begin{figure}[t]

\centering
\includegraphics[width=6.85cm]{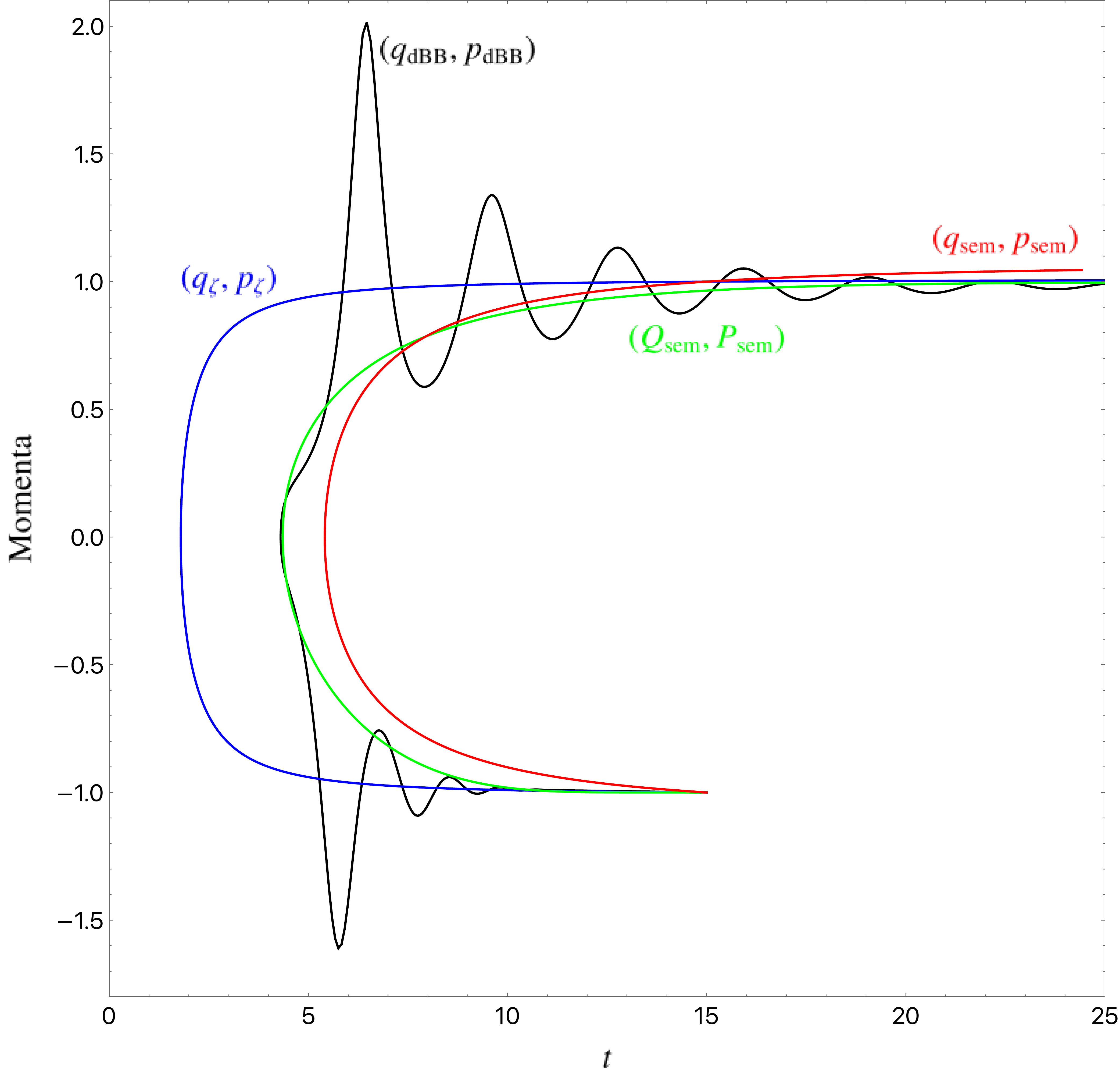}
\includegraphics[width=6.85cm]{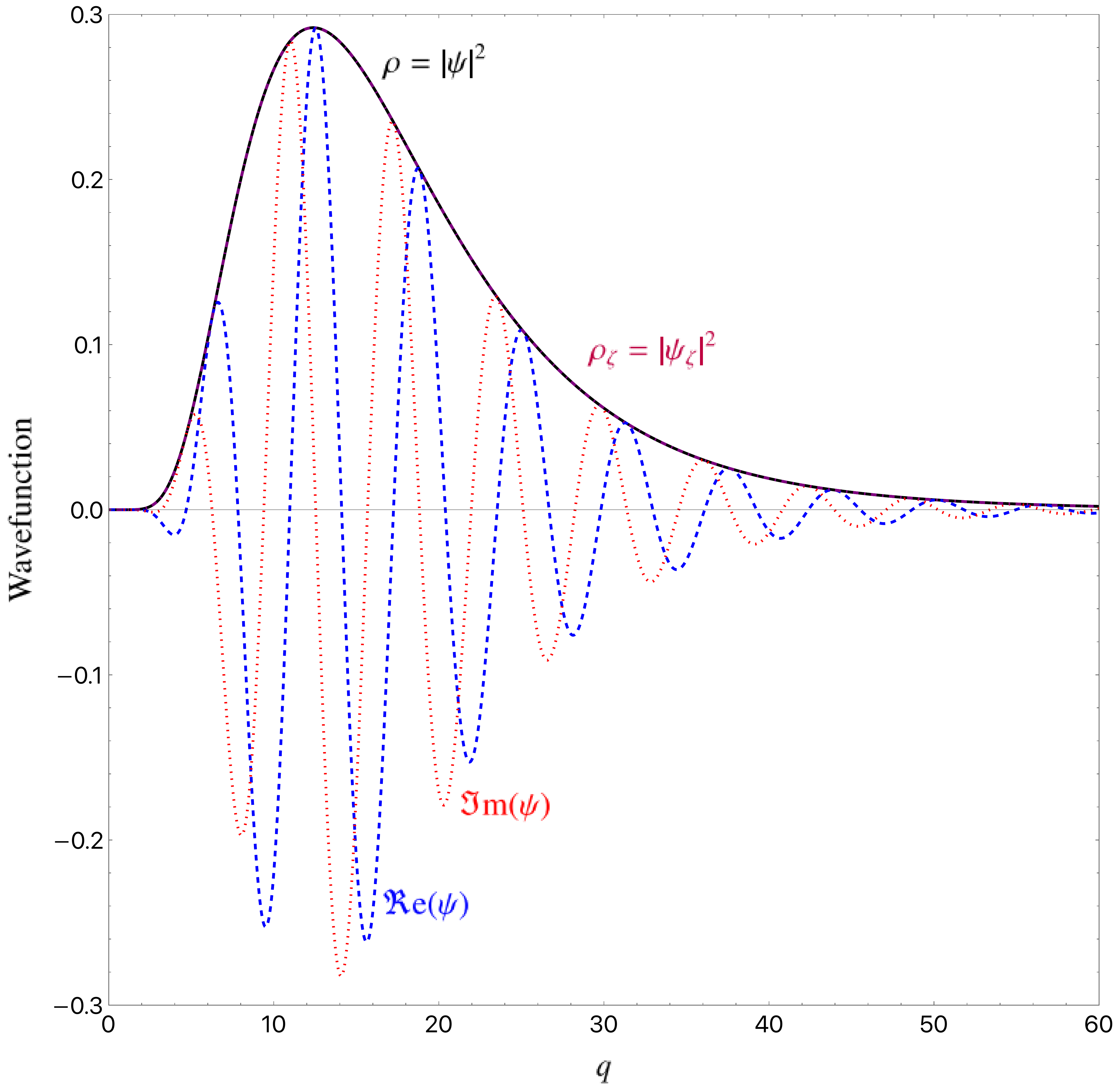}

\caption{(\textbf{Left}): Parametric phase space
trajectories built from the data from Figure~\ref{trajs}.
Except for the dBB case, all are well fitted (if not exactly
given) by $p^2 \propto q_\textsc{b}^{-2} -q^{-2}$, as
obtained from \eqref{solsem}. (\textbf{Right}): wavefunction
leading to the previous trajectories, at the initial time,
at which we assume a a coherent state. Subsequent evolution
is shown in Figure~\ref{psis}.}

\label{phase}
\end{figure}

Applying the coherent state method to the quantum
Hamiltonian \eqref{Hquantum} yields
\begin{equation}
H_\mathrm{sem}(q,p) = \frac12 p^2 + \frac{\mathcal{K}}{q^2},
\label{Hsem}
\end{equation}
in which $\mathcal{K} = \hbar^2\left[ \mathcal{K}_\xi
\rho_\zeta(1) + \sigma_\zeta(-2)\right]$. As above, the
coefficient $\mathcal{K}$ depends on the choice of fiducial
state $|\zeta\rangle$ and is, to a large extent, arbitrary.

Solving Equations~\eqref{eomsem} with \eqref{Hsem} yields
\begin{equation}
q_\zeta(t)=q_\textsc{b} \sqrt{1+(\omega t)^2} \quad
\hbox{and} \quad p_\zeta(t)=  \frac{q_\textsc{b}
    \omega^2 t}{\sqrt{1+(\omega t)^2}},
\label{solsem}
\end{equation}
where $q_\textsc{b} = \sqrt{\mathcal{K}/H_\text{sem}}$ and
$\omega = H_\text{sem}\sqrt{2/\mathcal{K}}$. It is
interesting to note that the solution \eqref{solsem} is
functionally the same as that obtained by using the operator
algebra $q_\mathrm{sem}(t)$ and $p_\mathrm{sem}(t)$, and
even though the parameters $q_\textsc{b}$ and $\omega$ in
both solutions differ in principle, they satisfy $\omega
q_\textsc{b} = \sqrt{2H}$ in both cases.

Finally, trajectories can be obtained in the quantum theory
of motion~\cite{Holland:1993ee} formulation of quantum
mechanics originally proposed by de Broglie in
1927~\cite{deBroglie:1927} and subsequently
formalised in more detail by Bohm in
1952 \cite{Bohm:1951xw,Bohm:1951xx}; we shall
accordingly refer in what follows to this formulation as the
de Broglie--Bohm (dBB) approach. Applied to quantum
gravity~\cite{Kiefer:2012ria}, it permits 
some relevant issues to be reformulated and, in some cases, solved
\cite{Pinto-Neto:2013toa}.

The basic idea stems from the eikonal approximation in the
classical wave theory of radiation for which light rays can
be obtained by merely following the gradients of the phase
of the wave. Similarly, in quantum mechanics, the
wavefunction is understood to represent an actual wave whose
phase gradient provides a means to calculate a trajectory.
In practice, for a Hamiltonian such as \eqref{Hquantum},
the Schr\"odinger equation reads $i\hbar\partial_t\psi = 
-\frac12 \partial^2_q \psi + V \psi$, which can be expanded,
setting $\psi(q,t) = \sqrt{\rho(q,t)} \exp[iS(q,t)/\hbar]$, into a
continuity equation
\begin{equation}
\frac{\partial\rho}{\partial t} + \frac{\partial}{\partial q} \left(
\rho \frac{\partial S}{\partial q} \right) =0,
\label{SchCont}
\end{equation}
naturally leading to the identification $\dot{q}_\dBB
=\partial_q S$, and a quantum-modified Hamilton--Jacobi
equation
\begin{equation}
\frac{\partial S}{\partial t} + \frac12 \left(
\frac{\partial S}{\partial q} \right)^2 +
V(q)+V_\textsc{q}=\frac{\partial S}{\partial t} + \frac12
\left( \frac{\partial S}{\partial q} \right)^2 + V(q)
\underbrace{-\frac{\hbar^2}{4\rho} \left[
\frac{\partial^2\rho}{\partial q^2} - \frac{1}{2\rho} \left(
\frac{\partial\rho}{\partial q} \right)^2
\right]}_{V_\textsc{q}} =0, \label{SchHJ}
\end{equation}
which confirms the above identification, while highlighting
a new potential adding to the original one. Appropriately
called the quantum potential, $V_\textsc{q}$, being built
out of the wavefunction solving the Schr\"odinger equation,
is in general a time-dependent potential.

\begin{figure}[t]
\centering
\includegraphics[width=6.85cm]{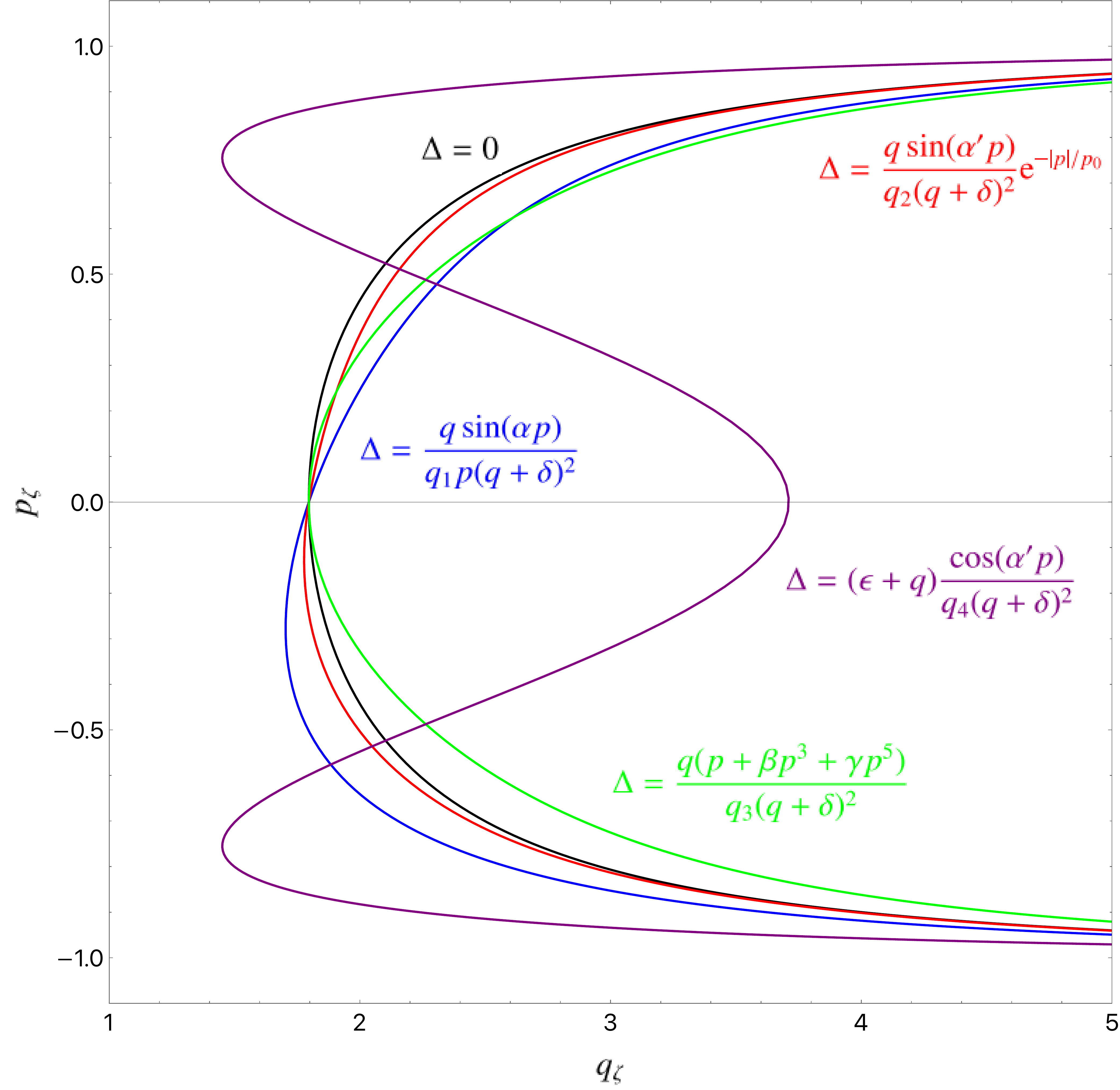}
\includegraphics[width=6.85cm]{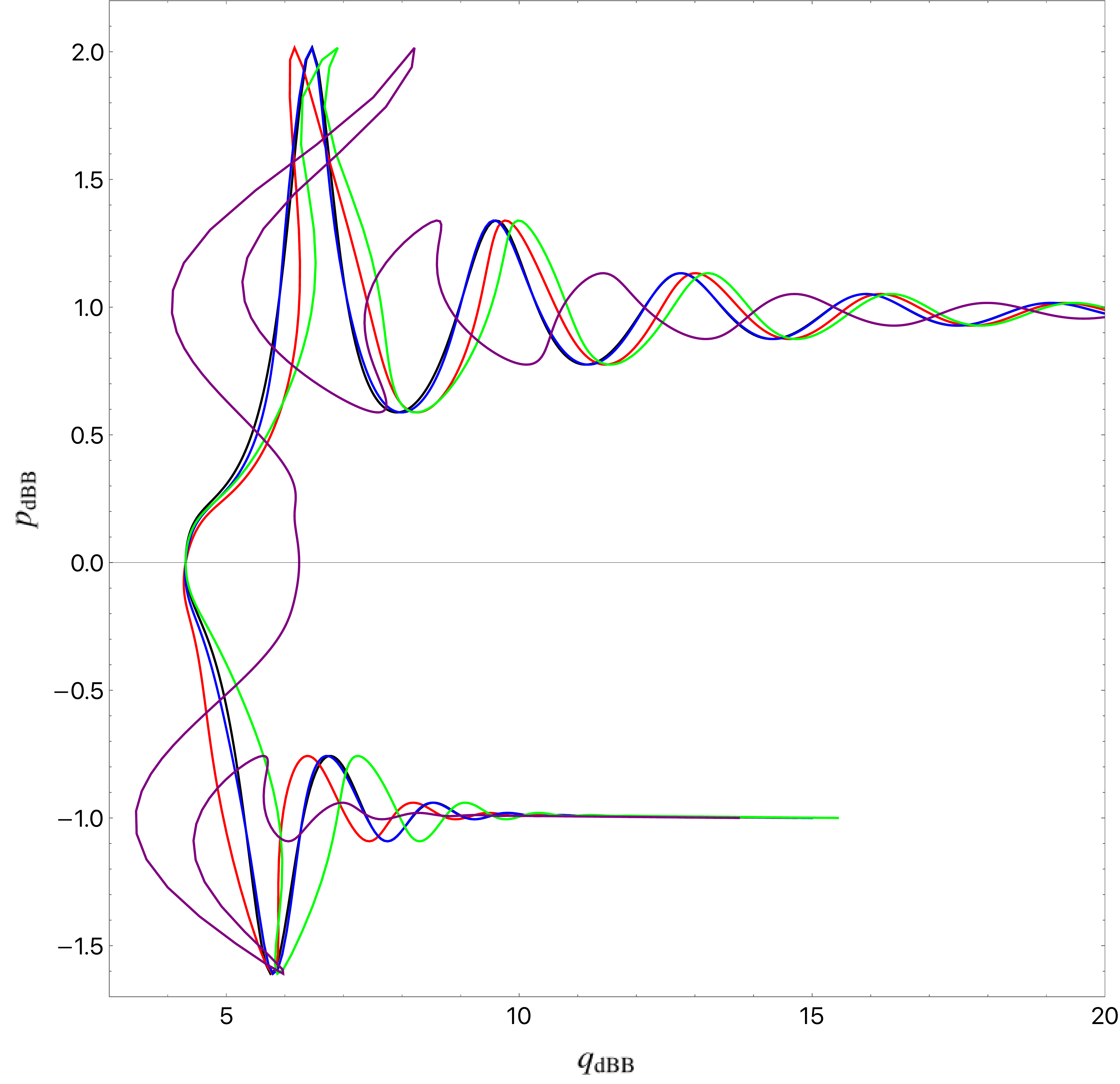}

\caption{Change of the phase space trajectories when a clock
transformation above \eqref{Delta} is applied to
Figure~\ref{phase}, with different time functions
$\Delta(q,p)$ as indicated in the figure. The numerical
parameters are set to $\alpha=2$, $\alpha'=3$, $\beta =
-0.63$, $\gamma=0.1$, $\epsilon=1$, $p_0=3/2$, $q_1=0.126$,
$q_2=0.4$, $q_3=0.04$, and $q_4=0.03$; they have been chosen
to yield visually important modifications of the
trajectories. Left panel: initial trajectory given by
Equation~\eqref{solsem}. Applying \eqref{QtPt} to this
trajectory yields the very same trajectory by definition.
Right panel: the dBB trajectory, initially having more
features, can be modified much more drastically.}

\label{clocks}
\end{figure}

With the identification $\dot{q}_\dBB=\partial_q S$, one
gets $\ddot{q}_\dBB = \dd (\partial_q S)/\dd t = \partial_t
(\partial_q S) + \partial^2_q S \dot{q}_\dBB$, so using the
identification again and the Hamilton--Jacobi equation
\eqref{SchHJ}, one finds\linebreak \mbox{$\ddot{q}_\dBB = - \partial_q (V+
V_\textsc{q})$}, i.e., a modified Newton equation that,
formally, can be derived from the time-dependent Hamiltonian
$H_\dBB = \frac12 p_\dBB^2 + V(q_\dBB) + V_\textsc{q}
(q_\dBB,t)$. These trajectories happen to be very different
from those derived above for various reasons. In particular,
the coherent state approximation leads to one and only one
trajectory $q_\zeta(t)$ once the initial coherent state
(including the fiducial state) is given.
Similarly, expectation values are unique for a given quantum
state, so that $q_\text{sem} (t)$ and $Q_\text{sem}(t)$
define one semiclassical or semiquantum approximation only,
which is entirely fixed by the parameters defining the
state, whereas $q_\dBB(t)$, stemming from a differential
equation, needs an initial value $q_\dBB(t_0)$ to be
evolved, and therefore there exists, for a given state, an
infinite number of acceptable trajectories. One could,
however, argue that for the coherent state trajectory, depending
on the choice of a particular fiducial state, there remains
some amount of ambiguity in this choice, permitting 
various families of such trajectories to be defined. In that sense,
the coherent state approximation and the dBB approach can
be compared.

Another crucial difference is that $q_\text{sem} (t)$,
$Q_\text{sem}(t)$ and $q_\zeta(t)$ represent \emph{
approximations} supposed to encode the underlying
quantum mechanical evolution of the wavefunction. The
trajectories $q_\dBB(t)$ are, by contrast, an extra degree
of freedom in the dBB formulation and thus exact solutions
of the equations of motion.

Let us consider beginning with the canonical quantisation
case, for which $\Ka \to 0$. In this case, our Bianchi I
vacuum model is formally equivalent, in the minisuperspace
limit, to that of a Friedmann universe filled with
radiation~\cite{AcaciodeBarros:1997gy}, and one finds that
there exists a wavefunction such that the $q_\dBB(t)$ has
the same functional dependence in time as $q_\zeta(t)$ in
Equation~\eqref{solsem}, except for the fact that the minimum
scale factor value is now given not only by the parameters
describing the wavefunction, but also depends on an initial
condition $q_\dBB(t_0)$. In that case, this comes from the
fact that the quantum potential happens to be $V_\textsc{q}
\propto q^{-2}$, so one naturally recovers the Hamiltonian
\eqref{Hsem}: one thus finds that all trajectories are
similar in shape.

The more relevant model in which $\Ka\not=0$ can also be
solved analytically under special conditions (see
Ref.~\cite{Malkiewicz:2019azw} for details and the solution
itself). Our choice in the present work was to assume an
initial wavefunction $\psi_\text{true}(q,t_\text{ini})$ in
the far past, with $q$ large, to be in a coherent state
$\psi_\zeta(q,t_\text{ini})$ (see the right panel of
Figure~\ref{phase}) and to evolve it with the Schr\"odinger
equation. Figure~\ref{psis} shows how $\psi (q,t) =
\psi_\text{true}(q,t)$ and then very rapidly departs from
$\psi_\zeta(q,t)$, although the expectation value
trajectories $q_\text{sem}$ and $Q_\text{sem}$ remain
similar (in shape, if not in actual values) to $q_\zeta$. As
it happens, as the wave packets move towards the origin
$q\to0$, $\psi$ starts oscillating, thus producing the
oscillations in the dBB trajectory $Q_\dBB$, while the
coherent state remains smooth at all times, being merely
squeezed close to the origin. It is interesting to note that
even though the wavefunctions differ drastically at the time
of the bounce, the relevant trajectories (except the dBB
one) are well described by \eqref{solsem}, although with
different parameters $q_\textsc{b}$ and $\omega$. We take
that as an indication that the coherent state approximation
is a valid one in most circumstances as long as one is only
interested in expectation values. Given the very significant
differences with the true wavefunction, however, it can be
assumed that higher order moments are {\sl not} well
approximated.

As a result, the trajectories defined through either expectation
values or coherent state approximation are invariant under
the clock transformation \eqref{QtPt}, contrary to the dBB ones.
However, as can be seen on Figure~\ref{clocks}, in which the
transformation stemming from the free particle Hamiltonian
is applied to the phase space trajectories, they do depend
on the choice of clock before quantisation. This is actually
even more true for the dBB case, for which these clock
transformations can lead to such tremendous modifications
of the space space trajectories that the actual
predictivity of the underlying theory becomes questionable.

\section{Conclusions}

We have reviewed the question of clock transformation and
trajectories in quantum cosmology by means of a simple
deparametrised and quantised Bianchi I model. The
Wheeler--DeWitt equation in this minisuperspace case
reduces to the Schr\"odinger equation of a free
particle or, depending on the quantisation scheme,
with a repulsive potential which can be studied using
standard techniques. The relevant degree of freedom,
from the point of view of cosmology, is the spatial
volume $q = V$, i.e., the cube of the scale factor
$a$, while the canonically conjugate momentum is
mostly given by the Hubble parameter.

Extending a previous work~\cite{Malkiewicz:2019azw} to
include dBB trajectories, we found very substantial
differences between those and their counterparts obtained by
some averaging processes. In the later case, all
trajectories stem from a semiclassical Hamiltonian and are
therefore invariant under the corresponding clock
transformation (although not for that corresponding to the
original classical theory). In the former case, however,
unless the wavefunction is restricted to belong to a very
special class (for which the coherent state approximation is
not valid), we found that the dBB trajectories depend in a
much more drastic way on the clock transformations,
rendering the ambiguity it stems from extremely serious, to
the point that the theory may  no longer even be predictive.
Calculating the spectrum of primordial perturbations, for
instance, involves the second time derivative of the scale
factor, and hence of our $q$, so that the choice of clock and
initial conditions can yield tremendously different
predictions. For semiclassical trajectories, on the other
hand, the choice is mostly irrelevant, and the resulting
perturbations might merely depend on a few parameters.

That said, it must be emphasised that the classical limit
is, in all cases (hence including dBB), well defined and
consistent, so there remains the possibility that whatever
dynamical quantity (e.g., perturbations) is evolved
through the full quantum phase might be unique. We postpone
such a discussion to a forthcoming work~\cite{Boldrin:2022}. 

\funding{This research was funded by the Polish National
Agency for Academic Exchange and Programme Hubert Curien
POLONIUM 2019 grant number 42657QJ.} 

\acknowledgments{The authors acknowledge many illuminating
discussions with H.~Bergeron, J.-P.~Gazeau and C.~Kiefer. }

\end{document}